\documentclass[12pt]{iopart}
\usepackage{iopams,graphicx,color}

\begin{document}
\def\be{\begin{equation}}
\def\ee{\end{equation}}
\def\bea{\begin{eqnarray}}
\def\eea{\end{eqnarray}}

\title{Stochastic Quantum Zeno by Large Deviation Theory}

\author{Stefano Gherardini$^{1,2,3}$, Shamik Gupta$^{4,5}$, Francesco Saverio Cataliotti$^{3,4}$, Augusto Smerzi$^{3,6}$, Filippo
Caruso$^{3,4}$, Stefano Ruffo$^{2,4}$}

\address{$^1$ \mbox{Department of Information Engineering, University of Florence,} via S. Marta 3, I-50139 Florence, Italy \\
$^2$ \mbox{CSDC, University of Florence, and INFN,} via G. Sansone 1, I-50019 Sesto Fiorentino, Italy \\
$^3$ \mbox{LENS, University of Florence,} via G. Sansone 1, I-50019 Sesto Fiorentino, Italy, and \mbox{QSTAR,} Largo E. Fermi 2, I-50125 Florence, Italy \\
$^4$ \mbox{Department of Physics and Astronomy, University of Florence,} via G. Sansone 1, I-50019 Sesto Fiorentino, Italy \\
$^5$ \mbox{Max Planck Institute for the Physics of Complex Systems}, Noethnitzer Stra\ss e 38, D-01187 Dresden, Germany \\
$^6$ \mbox{INO-CNR}, Largo E. Fermi 2, I-50125 Florence, Italy}

\begin{abstract}
Quantum measurements are crucial to observe the properties of a quantum system, which however unavoidably perturb its state and dynamics
in an irreversible way. Here we study the dynamics of a quantum system while being subject to a sequence of projective measurements applied at random times. In the case of independent and identically distributed intervals of time between consecutive measurements, we
analytically demonstrate that the survival probability of the system to remain in the projected state assumes a large-deviation (exponentially decaying) form in the limit of an infinite number of measurements. This allows us to estimate the typical value of the
survival probability, which can therefore be tuned by controlling the probability distribution of the random time intervals. Our analytical
results are numerically tested for Zeno-protected entangled states, which also demonstrates that the presence of disorder in the 
measurement sequence further enhances the survival probability when the Zeno limit is not reached (as it happens in experiments). Our studies provide a new tool for protecting and controlling the amount of quantum coherence in open complex quantum systems by means of tunable stochastic measurements.
\end{abstract}

\date{\today}


\maketitle

\section{Introduction}
A striking aspect of the dynamical evolution of quantum systems that distinguishes it from that of the classical ones is the strong influence on the
evolution caused by measurements performed on the system. Indeed, in the extreme case of a frequent enough series of measurements projecting the system back to the initial state,
its dynamical evolution gets completely frozen, i.e., the survival probability to remain in the initial state approaches unity in the limit of an infinite number of measurements.
This effect, known as the quantum Zeno effect (QZE), was first discussed in a seminal paper by Sudarshan and Misra in 1977 \cite{Misra1}, and can be understood
intuitively as resulting from the collapse of the wave function corresponding to the initial state due to the process of measurement.
It was later explored experimentally in systems of ions \cite{Itano1}, polarized photons \cite{Kwiat:1995}, cold atoms \cite{Fischer:2001}, and dilute Bose-Einstein condensed gases \cite{Streed:2006}. In noisy quantum systems, both the Zeno effect
and the acceleration due to an anti-Zeno effect \cite{kofman1,kofman2} have been demonstrated and also proposed for thermodynamical control of quantum systems \cite{Kurizki1}, and for quantum computation \cite{Lidar1}. 
The so-called quantum Zeno dynamics (QZD) that generalizes QZE is obtained when one applies frequent projective measurements onto a multi-dimensional Hilbert subspace \cite{Pascazio1}. In this case, the system does evolve away from
its initial state, but nevertheless remains confined in the subspace defined by the projection itself \cite{Pascazio2}. The QZD has been recently demonstrated in an experiment with a rubidium Bose-Einstein
condensate in a five-level Hilbert space \cite{Schafer1}. The evolution of physical observables can also be slowed down by frequent measurements (operator QZE) even while the quantum state changes randomly with
time \cite{Wang1}. Additionally, the Zeno phenomena have assumed particular relevance in applications owing to the possibility of quantum control, whereby specific quantum states (including entangled ones) may be
protected from decoherence by means of projective measurements \cite{Maniscalco1,Kim1}. Indeed, QZE is also a physical consequence of the statistical indistinguishability of neighboring quantum states \cite{Smerzi}.
Very recently, it has been shown that frequent observation of a small part of a quantum system turns its dynamics from very simple to an exponentially complex one \cite{Giovannetti1}, thereby paving the way for universal quantum computation.

While in its original QZE formulation, the measurement sequence is equally spaced in time, later treatments have considered the case of measurements randomly spaced in time (stochastic QZE) \cite{Shushin1}. In the latter case, the survival probability in the projected state becomes itself a random variable that takes on different values corresponding to different realizations of the measurement sequence. In particular, one would expect that the mean of the survival probability obtained by considering an average over
a large (ideally infinite) number of realizations of the measurement sequence leads to the result obtained for an evenly spaced sequence, under some constraints (e.g., the mean of the time interval between consecutive measurements is finite). An
interesting question, relevant both theoretically and experimentally, naturally emerges: Is it possible to have realizations of the measurement sequence that give values of the survival probability significantly deviated
from the mean? How typical/atypical are those realizations? Are there ways to quantify the probability measures of such realizations? These questions assume particular
importance in devising experimental protocols that on demand may slow down or speed up efficiently the transitions of a quantum system between its possible states.

In this work, by exploiting tools from probability theory, we propose a framework that allows an effective addressal of the questions posed above. In particular, we adapt the well-established theory of large deviations (LD) to quantify the dependence of the survival probability
on the realization of the measurement sequence, in the case of independent and identically distributed (i.i.d.) time intervals between consecutive measurements. Our goal is two fold: 1) adapt and apply the LD theory to discuss the QZE by transferring
tools and ideas from classical probability theory to the arena of quantum Zeno phenomena, 2) analytically predict the corresponding survival probability and exploit it for a new type of control based on the stochastic features of the applied measurements.

The LD theory concerns the asymptotic exponential decay of probabilities associated with large fluctuations of stochastic variables. Originally developed in the realm of probability theory \cite{Varadhan:1984,Ellis1,Dembo1,Touchette1}, an increasing
interest in the last years has led to several studies of large deviations in both classical and quantum systems. In the latter case, the LD formalism has been discussed in the context of quantum gases \cite{Gallavotti1}, quantum spin systems \cite{Netovcny1}, quantum
information theory \cite{Ahlswede1}, among others. An interesting recent application pursued in Refs. \cite{Garrahan1,Budini1,Garrahan3,Garrahan4} has invoked the LD theory to develop a thermodynamic formalism for the study of quantum jump trajectories \cite{Plenio1,Hegerfeldt1} of open quantum systems \cite{Davies1,Petruccione1,Caruso1}. Indeed, they have addressed thermodynamic issues for quantum systems, e.g., quantum stochastic thermodynamics for the quantum trajectories of a continuously monitored
forced harmonic oscillator coupled to a thermal reservoir \cite{Horowitz1}, the work statistics in a driven two-level system coupled to a heat bath \cite{Pekola1}, and stochastic thermodynamics of a quantum heat engine \cite{Campisi1}.

Here, we consider a general quantum system with unitary dynamical evolution that is subject to a sequence of measurements projecting it into a fixed (initial) state. These measurements are separated by time intervals that are i.i.d. random variables. We analytically show that in the limit of a large number $m$ of measurements, the distribution of the survival probability to remain in the initial state assumes a large-deviation form, namely, a profile decaying exponentially in $m$ with a positive multiplying factor, the so-called rate function, which is a function
only of the survival probability. The value at which the function attains its minimum gives out of all possible outcomes the most probable or the typical value of the survival probability. Our analytical results are supported by numerical studies in
the case of Zeno-protected entangled states. They show that the presence of disorder in the sequence of time intervals between consecutive measurements is deleterious in reaching the Zeno limit.
Nevertheless, the disorder does enhance the survival probability when the latter is not exactly one, which, interestingly enough, corresponds to the typical experimental situation.
 
The layout of the paper is as follows. In Section \ref{model}, we introduce our framework applied to a generic quantum system with unitary dynamics and subjected to a sequence of projective measurements performed at random times. We then discuss in Section
\ref{survival-probability} the statistics of the survival probability of the system to remain in an initial pure state. In Section \ref{LD}, we consider the case of a $d$-dimensional Bernoulli probability distribution for the
time intervals between consecutive measurements, and analyze the statistics of the survival probability in the limit of a large number of measurements by means of the LD formalism. We also generalize our results to the case of a continuous probability distribution. 
In Section \ref{numerics}, we confirm our analytical results by numerical studies on Zeno-protected entangled states for a generic three-level quantum system. In Section \ref{QZE}, we discuss how to analytically recover the exact quantum Zeno limit.
Conclusions and outlook follow in Section \ref{conclusions}. Some of the technical details of our analysis are provided in the four Appendices.

\section{The Model}
\label{model}
Consider a quantum mechanical system described by a finite-dimensional Hilbert space $\mathcal{H}$, which may be taken to be a direct sum of $r$ orthogonal subspaces $\mathcal{H}^{(k)}$, as $\mathcal{H}=\bigoplus_{k=1}^{r}\mathcal{H}^{(k)}$. We assign to each
subspace a projection operator $\textbf{P}^{(k)}$, i.e., $\textbf{P}^{(k)}\mathcal{H}=\mathcal{H}^{(k)}$. An initial quantum state described by the density matrix $\rho_0$ undergoes a unitary dynamics to evolve in time $t$ to
$\exp(-i\textbf{H}t)\rho_{0}\exp(i\textbf{H}t)$, where $\textbf{H}$ is a generic system Hamiltonian that in general commutes with the projectors $\textbf{P}^{(k)}$. In this work, we set the reduced Planck's constant $\hbar$ to unity.

In our model, starting with a $\rho_{0}$ that belongs to one of the subspaces, say subspace $\bar{r}\in 1,2\ldots,r$, so that $\rho_{0}=\textbf{P}^{(\bar{r})}\rho_{0}\textbf{P}^{(\bar{r})}$ and
$\rm{Tr}[\rho_{0}\textbf{P}^{(\bar{r})}]=1$, we subject the system to an arbitrary but fixed number $m$ of consecutive measurements separated by time intervals $\mu_j;~\mu_j >0$, with $j=1,\ldots,m$. During each interval $\mu_j$,
the system follows a unitary evolution described by the Hamiltonian $\textbf{H}$, while the measurement corresponds to applying the projection operator $\textbf{P}^{(\bar{r})}$. We take the $\mu_j$'s to be
independent and identically distributed (i.i.d.) random variables sampled from a given distribution $p(\mu)$, with the normalization $\int d\mu~p(\mu)=1$. We assume that $p(\mu)$ has
a finite mean, denoted by $\overline{\mu}$. For simplicity, in the following we represent $\textbf{P}^{(\bar{r})}$ and
$\mathcal{H}^{(\bar{r})}$ by $\textbf{P}$ and $\mathcal{H}_{P}$, respectively. The (unnormalized) density matrix at the end of evolution for a total time
\be\label{T-defn}
\mathcal{T}\equiv \sum_{j=1}^m\mu_j,
\ee
corresponding
to a given realization of the measurement sequence
$\{\mu_j\}\equiv\{\mu_j;~j=1,2,\ldots,m\}$, is
given by
\bea
\textbf{W}_m(\{\mu_j\})&\equiv&\left(\textbf{P}\textbf{U}_m\right)\ldots\left(\textbf{P}\textbf{U}_1\right)\rho_{0}
\left(\textbf{P}\textbf{U}_1\right)^\dagger\ldots\left(\textbf{P}\textbf{U}_m\right)^\dagger
\nonumber \\
&=&\textbf{R}_{m}^{\dagger}(\{\mu_j\})\rho_{0}\textbf{R}_{m}(\{\mu_j\}),
\label{eq:Wm}
\eea
where we have defined
\begin{equation}\label{4}
\textbf{R}_m(\{\mu_j\})\equiv\prod_{j=1}^{m}\textbf{P}\textbf{U}_j\textbf{P},
\end{equation}
and
\be\label{U-defn}
\textbf{U}_j\equiv\exp\left(-i\textbf{H}\mu_j\right).
\ee
To obtain Eq. (\ref{eq:Wm}), we have used $\textbf{P}^\dagger=\textbf{P}$, $\rho_{0}=\textbf{P}\rho_{0}\textbf{P}$, and $\textbf{P}^2=\textbf{P}$.
Note that $\mathcal{T}$ is a random variable that depends on the realization of the sequence $\{\mu_j\}$.
The survival probability, namely, the probability that the system belongs to the subspace $\mathcal{H}_{P}$ at the end of the
evolution, is given by
\bea\label{3}
\mathcal{P}(\{\mu_j\})\equiv\rm{Tr}\left[\textbf{W}_{m}(\{\mu_j\})\right]=\rm{Tr}\left[\textbf{R}_{m}^{\dagger}(\{\mu_j\})\rho_{0}\textbf{R}_{m}(\{\mu_j\})\right],
\eea
while the final (normalized) density matrix is
\begin{equation}\label{5}
\rho(\{\mu_j\})=\frac{\textbf{R}_{m}^{\dagger}(\{\mu_j\})\rho_{0}\textbf{R}_{m}(\{\mu_j\})}{\mathcal{P}(\{\mu_j\})}.
\end{equation}
Note that the survival probability $\mathcal{P}(\{\mu_j\})$ depends on
the system Hamiltonian $\textbf{H}$, the initial density matrix
$\rho_0$, and also on the probability distribution $p(\mu)$.

\section{Statistics of the survival probability}
\label{survival-probability}
In this section, we obtain the distribution of the survival probability $\mathcal{P}(\{\mu_j\})$ with respect to different
realizations of the sequence $\{\mu_j\}$. We suppose that the system is initially in a pure state $|\psi_{0}\rangle$
belonging to $\mathcal{H}_{P}$, so that $\rho_{0}=|\psi_{0}\rangle\langle\psi_{0}|$, and assume that the projection operator
is $\textbf{P}\equiv|\psi_{0}\rangle\langle\psi_{0}|$. In this way, starting with a pure state, the system evolves according to the
following repetitive sequence of events: unitary evolution for a random interval, followed by a measurement that projects the evolved state into
the initial state. Note that our analysis can be easily generalized to the case of initially mixed states and for other choices of the projection operator.

The survival probability $\mathcal{P}(\{\mu_j\})$ can be evaluated by using Eq. (\ref{3}) to get 
\begin{equation}
\mathcal{P}(\{\mu_j\})=\prod_{j=1}^{m}q(\mu_j), \label{6}
\end{equation}
where we have defined the probability $q(\mu_j)$ as
\begin{equation}\label{7}
q(\mu_j)\equiv\left|\langle\psi_{0}|\textbf{U}_j|\psi_{0}\rangle\right|^{2},
\end{equation}
which takes on different values depending on the random numbers $\mu_j$. Note that being a probability, possible values of $q(\mu)$ lie in the range $0 < q(\mu) \le 1$. The distribution of $q(\mu_j)$ is obtained as
\begin{equation}\label{8}
P\left(q(\mu_j)\right)=p(\mu_j)\left|\frac{d\mu_j}{dq(\mu_j)}\right|,
\end{equation}
where Eq. (\ref{7}) gives
\begin{equation}\label{9}
\left|\frac{dq(\mu_{j})}{d\mu_j}\right|=2\left|\langle\psi_{0}|\textbf{H}\textbf{U}_j|\psi_{0}\rangle\right|.
\end{equation}
From Eq. (\ref{6}), one then derives the distribution of $\mathcal{P}$ as
\be\label{9a}
P\left(\mathcal{P}\right)=\left[\prod_{j=1}^{m}\int d\mu_j~ p(\mu_j)\right]\delta\left(\prod_{j=1}^{m}q(\mu_{j})-
\mathcal{P}\right).
\ee
In particular, one may be interested in the average value of the survival probability, where the average corresponds to repeating a large
number of times the protocol of $m$ consecutive measurements interspersed with unitary dynamics for random intervals $\mu_j$. One gets:
\begin{equation}\label{9b}
\langle
\mathcal{P}\rangle=\prod_{j=1}^{m}\int d\mu_j~ p(\mu_j)q(\mu_{j}).
\end{equation}
Here and in the following, we will use angular brackets to denote averaging with respect to different realizations of the sequence
$\{\mu_j\}$. Additionally, let us note that writing $q(\mu)$ as 
\be\label{qmu-expansion0}
q(\mu)=1-\delta(\mu);~0 \le \delta(\mu) < 1,
\ee
we have
\be\label{qmu-expansion1}
\delta(\mu)=\left|\sum_{k=1}^\infty \frac{(-i\mu)^k}{k!}\langle \textbf{H}^k\rangle\right|^2,
\ee
with
\be
\langle \textbf{H}^k\rangle \equiv \langle \psi_0|\textbf{H}^k|\psi_0\rangle;~k=0,1,2,\ldots.
\ee
In particular, considering $\mu \ll 1$, one has, to leading order in $\mu^2$, the result
\be\label{qmu-expansion2}
\delta(\mu)=\frac{\mu^2}{\tau_Z^2},
\ee
where $\tau_Z$ is the so-called Zeno-time \cite{Pascazio2}:
\bea
\tau_Z^{-2}&\equiv \Delta^{2}\textbf{H}, \\
\Delta^{2}\textbf{H} &\equiv \langle\textbf{H}^2\rangle-\langle\textbf{H}\rangle^2.
\eea
\section{Large deviation formalism for the survival probability}
\label{LD}
Let us now employ the LD formalism to discuss the statistics of the survival probability $\mathcal{P}(\{\mu_j\})$ in the limit $m \to
\infty$. In this limit, Eq. (\ref{T-defn}) gives
\be\label{Tavg-1}
\langle \mathcal{T}\rangle=m\overline{\mu},
\ee
where we have used the fact that the $\mu_j$'s are i.i.d. random variables and $\overline{\mu}$ is a finite number.

To proceed further, let us consider $p(\mu)$ to be a $d$-dimensional Bernoulli distribution, namely, $\mu$ takes on $d$ possible discrete values $\mu^{(1)},\mu^{(2)},\ldots,\mu^{(d)}$ with corresponding probabilities $p^{(1)},p^{(2)},\ldots,p^{(d)}$, such that $\sum_{\alpha=1}^d p^{(\alpha)}=1$. The average value of the survival probability is then obtained by using Eq. (\ref{9b}) as
\begin{equation}\label{P-avg}
\langle\mathcal{P}\rangle=\exp\Big(m\ln\sum_{\alpha=1}^d p^{(\alpha)}q(\mu^{(\alpha)})\Big).
\end{equation}

In order to introduce the LD formalism for the survival probability, first consider
\begin{equation}\label{10}
\mathcal{L}(\{\mu_j\})\equiv\ln\left(\mathcal{P}(\{\mu_j\})\right)=\sum_{\alpha=1}^{d}n_{\alpha}\ln q(\mu^{(\alpha)}),
\end{equation}
where
$n_{\alpha}$ is the number of times $\mu^{(\alpha)}$ occurs in the sequence $\{\mu_j\}$. Noting that $\mathcal{L}(\{\mu_j\})$ is a sum of
i.i.d. random variables, its probability distribution is given by
\begin{eqnarray}\label{11}
P(\mathcal{L})&=&\sum_{n_1,n_2,\ldots n_d:~\sum_{\alpha=1}^d
n_\alpha=m}\frac{m!}{n_{1}!n_{2}!\ldots n_{d}!}(p^{(1)})^{n_1}\ldots (p^{(d)})^{n_d} \nonumber \\
&&\times \delta\left(\sum_{\alpha=1}^{d}n_{\alpha}\ln q(\mu^{(\alpha)})-\mathcal{L}\right)\nonumber \\
&=&\frac{m!}{n^\prime_{1}!n^\prime_{2}!\ldots
n^\prime_{d}!}\prod_{\alpha=1}^{d}(p^{(\alpha)})^{n^\prime_\alpha}
\label{eq1},
\end{eqnarray}
where, as indicated, the summation in the first equality is over all possible values of $n_1,n_2,\ldots,n_d$ subject to the constrain $\sum_{\alpha=1}^d
n_\alpha=m$. In the second equality, $n^\prime_\alpha$'s are such that
\begin{eqnarray}
&&\sum_{\alpha=1}^d n^\prime_{\alpha}=m,\label{12} \\
&&\sum_{\alpha=1}^{d}n^\prime_{\alpha}\ln q(\mu^{(\alpha)})=\mathcal{L}.\label{12a}
\end{eqnarray}

Starting from Eq. (\ref{eq1}), and considering the limit $m \to \infty$,
a straightforward manipulation (see \ref{app1}) leads to the following
LD form for the probability distribution
$P\left(\mathcal{L}/m\right)$, as
\begin{equation}\label{15}
P\left(\mathcal{L}/m\right)\approx
\exp\Big(-mI\left(\mathcal{L}/m\right)\Big),
\end{equation}
where the function $I(x)$, i.e., the so-called rate function \cite{Touchette1},
is given by
\begin{eqnarray}
&&I\left(x\right)=\sum_{\alpha=1}^{d}f(\mu^{(\alpha)})\ln\left(\frac{f(\mu^{(\alpha)})}{p^{(\alpha)}}\right);\label{16}
\\
&&f(\mu^{(\alpha)})=\frac{\ln q(\mu^{(d)})-x}{(d-1)\Big[\ln q(\mu^{(d)})-\ln q(\mu^{(\alpha)})\Big]};~\alpha=1,\ldots,(d-1)\label{17}, \\
&&f(\mu^{(d)})=1-\sum_{\alpha=1}^{d-1}f(\mu^{(\alpha)})\label{17b}.
\end{eqnarray}
The approximate symbol $\approx$ in Eq. (\ref{15}) stands for the fact that there are subdominant $m$-dependent factors on the right hand side
of the equation. An alternative form to Eq. (\ref{15}) that involves an
exact equality and can be considered as the equation defining the
function $I(x)$ is
\begin{equation}
\lim_{m\to \infty}
-\frac{1}{m}P\left(\mathcal{L}/m\right)=I\left(\mathcal{L}/m\right).
\end{equation}

It is evident from Eq. (\ref{16}) that the rate function $I\left(x\right)$ is the relative entropy or the Kullback-Leibler distance between the set of
probabilities $\{f(\mu^{(\alpha)})\}$ and the set $\{p^{(\alpha)}\}$; it
has the property to be positive and convex, with a single non-trivial
minimum \cite{Cover1}. Equation (\ref{15}) implies that the value at which the function $I(\mathcal{L}/m)$ is minimized corresponds to the most probable value $\mathcal{L}^{\star}$ of
$\mathcal{L}$ as $m \to \infty$. Using $\partial
I(\mathcal{L}/m)/\partial \ln q(\mu^{(\alpha)})|_{\mathcal{L}=\mathcal{L}^\star}=0;~
\alpha=1,\ldots,d$, we get (see \ref{app2})
\begin{equation}\label{21}
\mathcal{L}^\star=m\sum_{\alpha=1}^{d} p^{(\alpha)}\ln q(\mu^{(\alpha)}).
\end{equation}

As for the distribution of the survival probability, one may obtain a
LD form for it in the following way:
\begin{eqnarray}\label{surv-prob-distr}
P(\mathcal{P}) &=\int d\mathcal{L}~P(\mathcal{L})\delta(\mathcal{L}-\mathcal{\ln P})\nonumber \\
&=\int d(\mathcal{L}/m)~P(\mathcal{L}/m)\delta(\mathcal{L}/m-\mathcal{\ln P})\nonumber \\
 & \approx\int d(\mathcal{L}/m)\exp\left(-mI(\mathcal{L}/m)\right)\delta(\mathcal{L}/m-\mathcal{\ln P}))\nonumber \\
 & \approx\exp\Big(-m ~{\rm
 min}_{\mathcal{L}:\mathcal{L}=m\ln\mathcal{P}}I(\mathcal{L}/m)\Big),
\end{eqnarray}
where in the third step we have considered large $m$ and have used
Eq. (\ref{15}), while in the last step we have used the saddle point method to evaluate the integral. We thus have
\begin{eqnarray}\label{18}
\lim_{m\to \infty}
-\ln(P(\mathcal{P}))/m=J(\mathcal{P});~J(\mathcal{P})\equiv{\rm min}_{\mathcal{L}:\mathcal{L}=m\ln\mathcal{P}}I(\mathcal{L}/m).
\end{eqnarray}
The value at which $J(\mathcal{P})$ takes on its minimum value gives the most probable value of the
survival probability in the limit $m \to \infty$, which may also be obtained by utilizing the relationship between $\mathcal{L}$ and $\mathcal{P}$; one gets
\begin{equation}\label{20b}
\mathcal{P}^{\star}=\exp\Big(m\sum_{\alpha=1}^{d} p^{(\alpha)}\ln q(\mu^{(\alpha)})\Big),
\end{equation}
which may be compared with the average value (\ref{P-avg}). In other words, while the average value $\langle \mathcal{P} \rangle$ is determined by the logarithm of the averaged $q(\mu^{(\alpha)})$, 
the most probable value $\mathcal{P}^{\star}$ is given by the average performed on the logarithm of $q(\mu^{(\alpha)})$. The latter is the so-called log-average or the geometric mean of the quantity $q(\mu^{(\alpha)})$
with respect to the $\mu$-distribution. A straightforward generalization of Eq. (\ref{20b}) for a generic (continuous) $\mu$-distribution is 
\be\label{PG-defn}
\mathcal{P}^\star=\exp\Big(m\int d\mu~ p(\mu)\ln q(\mu)\Big),
\ee
while that for the average reads
\be\label{P-avg1}
\langle \mathcal{P} \rangle=\exp\Big(m \ln \int d\mu~ p(\mu)q(\mu)\Big).
\ee
Using the so-called Jensen's inequality, namely, $\langle \exp
(x)\rangle \ge \exp(\langle x\rangle)$, it immediately follows that
\be
\langle \mathcal{P} \rangle \ge \mathcal{P}^\star,
\ee
with the equality holding only when no randomness in $\mu$ (that is, only a single value of $\mu$ exists) is considered. The difference between $\mathcal{P}^\star$ and
$\langle \mathcal{P}\rangle$ can be estimated in the following way in experiments. First, we perform a large number $m$ of projective
measurements on our quantum system. The value of the survival probability to remain in the initial state that is measured in a single
experimental run will very likely be close to $\mathcal{P}^\star$, with deviations that decrease fast with increasing $m$. On the other hand, averaging the survival probability over a large (ideally infinite) number of experimental runs will yield $\langle \mathcal{P}\rangle$.

All the derivations above were based on the assumption of a fixed number $m$ of measurements, so that the total time interval $\mathcal{T}$ -- see Eq.
(\ref{T-defn}) -- is a quantity fluctuating between different realizations of the measurement sequence. To obtain the LD formalism, we eventually let $m$
approach infinity, which in turn leads to an infinite $\langle\mathcal{T} \rangle$ (unless $\overline{\mu} \to 0$) -- see Eq.
(\ref{Tavg-1}). We now consider the situation where we keep the total
time $\mathcal{T}$ fixed, and let $m$ fluctuate between realizations of the measurement
sequence. In this case, in contrast to Eq. (\ref{11}), we have the joint probability distribution
\begin{eqnarray}\label{32}
&&P(\mathcal{L},\mathcal{T})=\sum_{m}\sum_{n_1,n_2,\ldots,n_d:~\sum_{\alpha=1}^d
n_\alpha=m}\frac{m!}{n_{1}!\ldots n_{d}!}\prod_{\alpha=1}^{d}\left(p^{(\alpha)}\right)^{n_{\alpha}}\nonumber \\
&&\times\delta\left(\sum_{\alpha=1}^{d}n_{\alpha}\ln
q(\mu^{(\alpha)})-\mathcal{L}\right)\delta\left(\sum_{\alpha=1}^{d}n_{\alpha}\mu^{(\alpha)}-\mathcal{T}\right).
\end{eqnarray}
We thus have to find the set of $n_\alpha$'s, which we now refer to as
$n^\prime_\alpha$'s, such that the following conditions are
satisfied:
\begin{eqnarray}
&&\sum_{\alpha=1}^d n^\prime_{\alpha}=m, \label{cons1}\\
&&\sum_{\alpha=1}^{d}n^\prime_{\alpha}\ln
q(\mu^{(\alpha)})=\mathcal{L}, \label{cons2} \\
&&\sum_{\alpha=1}^{d}n_{\alpha}'\mu^{(\alpha)}=\mathcal{T}.
\label{cons3}
\end{eqnarray}
The above equations have a unique solution only for $d=2$, that is, when one
has a Bernoulli distribution. In this case, the solutions satisfy
\begin{equation}\label{33}
\frac{\mathcal{T}-m\mu^{(2)}}{\mu^{(2)}-\mu^{(1)}}=\frac{\mathcal{L}-m\ln
q(\mu^{(2)})}{\ln q(\mu^{(2)})-\ln q(\mu^{(1)})},
\end{equation}
which may be solved for $m$, for given values of $\mathcal{L}$ and
$\mathcal{T}$ and then used in Eq. (\ref{32}) to determine
$P(\mathcal{L},\mathcal{T})$.
In the limit $m\rightarrow\infty$, provided the mean
$\overline{\mu}$ of $p(\mu)$ exits, Eq. (\ref{T-defn}) together with the
law of large numbers\footnote{The law of large numbers states that the sum of a large
number $N$ of i.i.d. random variables, when scaled by $N$, tends to the mean of
the underlying identical distribution with probability one as $N$
approaches infinity \cite{Feller1}.} gives:
\be\label{T-defn1}
\mathcal{T}=m\overline{\mu}.
\ee
In this case, for every $d$, one obtains an LD form for
$P(\mathcal{L},\mathcal{T})$ (see \ref{app3}):
\begin{eqnarray}\label{LD-PLT}
P(\mathcal{L},\mathcal{T})&\approx&\exp\Big(-m\mathcal{I}\left(\frac{\mathcal{L}}{m},\frac{\mathcal{T}}{m}\right)\Big),
\end{eqnarray}
where
\begin{eqnarray}\label{40}
&&\mathcal{I}\left(x,y\right)=\sum_{\alpha=1}^{d}g(\mu^{(\alpha)})\ln\left(\frac{g(\mu^{(\alpha)})}{p^{(\alpha)}}\right),
\\
&&g(\mu^{(\alpha)})=\frac{\mu^{(d)}(m\ln q\left(\mu^{(d)}\right)-x)}
{\left(\ln
q\left(\mu^{(d)}\right)-x\right)(\mu^{(d)}-\mu^{(\alpha)})+y(d-1)\ln\left(q\left(\mu^{(d)}\right)/q\left(\mu^{(\alpha)}\right)\right)};\nonumber
\\
&&~~~~~~~~~~~~~~~~~~~~~~~~~~~~~~~~~~~~~~~~~~~~~~~~~~~\alpha=1,\ldots,(d-1), \\
&&g(\mu^{(d)})=1-\sum_{\alpha=1}^{d-1}g(\mu^{(\alpha)}).
\end{eqnarray}
The rate function (\ref{40}) is related to the rate function (\ref{16})
as follows:
\begin{equation}\label{42}
I(\mathcal{L}/m)={\rm min}_{\mathcal{T}/m: \mathcal{T}=\langle
\mathcal{T}\rangle}\mathcal{I}\left(\frac{\mathcal{L}}{m},\frac{\mathcal{T}}{m}\right),
\end{equation}
and similarly to Eq. (\ref{surv-prob-distr}), one has
\begin{eqnarray}
P(\mathcal{P},\mathcal{T})&\approx\exp\left(-m
\mathcal{J}(\mathcal{P},\mathcal{T}/m)\right),
\end{eqnarray}
where
\begin{eqnarray}
\mathcal{J}(\mathcal{P},\mathcal{T}/m)\equiv{\rm
min}_{\mathcal{L}:\mathcal{L}=m\ln\mathcal{P}}\mathcal{I}(\mathcal{L}/m,\mathcal{T}/m).
\end{eqnarray}
{As in} Eq. (\ref{PG-defn}), the most probable value of the survival probability for a continuous $\mu$-distribution is given by
\be
\mathcal{P}^\star(\mathcal{T})=\exp\Big(m\int
d\mu~ p(\mu)\ln g(\mu)\Big).
\ee

\section{Numerical results}
\label{numerics}
In this section, in order to test our analytical results, we numerically simulate the dynamical evolution of a generic $n$-level quantum system governed by the following Hamiltonian
\begin{equation}\label{H}
H=\sum_{j=1}^{n}\omega_{j}|j\rangle\langle j|+\sum_{j=1}^{n-1}\Omega\left(|j\rangle\langle j+1| + |j+1\rangle\langle j|\right).
\end{equation}
Here, $|j\rangle \equiv |0 \dots 1 \dots 0 \rangle$, with $1$ in the $j$-th place and $0$ otherwise, denotes the state for the $j$-th level, ($j=1,\dots,n$), with the corresponding energy $\omega_{j}$, while $\Omega$ is the coupling rate between nearest-neighbor levels. For simplicity, we take $n=3$, $\Omega=2\pi f$, with $f=100$ kHz, and $\omega_{j}=2\pi f_{j}$, with $f_{1}=30$ kHz, $f_{2}=20$ kHz and $f_{3}=10$ kHz.
We choose the initial state $|\psi_0\rangle$ to be the entangled (with respect to the bipartition $1|23$) pure state
\be\label{initial-state}
|\psi_{0}\rangle \equiv \frac{1}{\sqrt{2}}(|100\rangle+|001\rangle).
\ee
\begin{figure}[b]
\centering
\includegraphics[width=105mm]{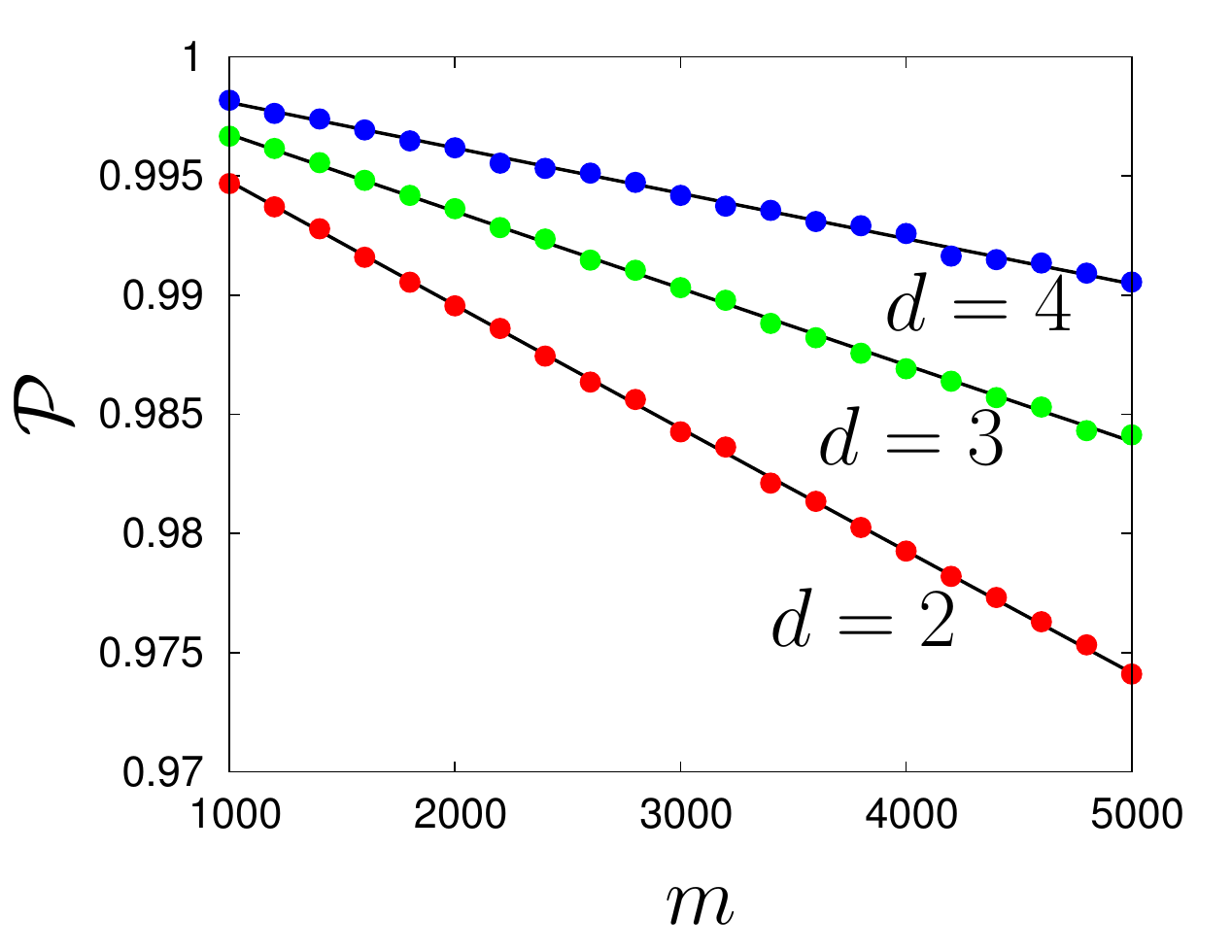}
\caption{Survival probability $\mathcal{P}$ as a function of the number of
measurements $m$ for a $d$-dimensional Bernoulli
distribution for the $\mu_j$'s, with $d=2,3,4$. Specifically, we have
chosen for $d=4$ the values $p^{(1)}=0.3, ~p^{(2)}=0.2, ~p^{(3)}=0.05,
~p^{(4)}=0.45$, and
$\mu^{(1)}=\mu_0,~\mu^{(2)}=3\mu_0,~\mu^{(3)}=2\mu_0,~\mu^{(4)}=0.5\mu_0$, with $\mu_0=1$ ns.
For $d=3$, the values are $p^{(1)}=0.3, ~p^{(2)}=0.2, ~p^{(3)}=0.5$, and
$\mu^{(1)}=\mu_0,~\mu^{(2)}=3\mu_0,~\mu^{(3)}=2\mu_0$, while for $d=2$,
we have taken $p^{(1)}=0.3, ~p^{(2)}=0.7$, and
$\mu^{(1)}=\mu_0,~\mu^{(2)}=3\mu_0$. Here, the points denote the values obtained by
evaluating Eq. (\ref{3}) numerically for a typical realization of
the measurement sequence $\{\mu_j\}$, while the lines denote the
asymptotic most probable values obtained by using Eq.
(\ref{20b}).}
\label{dnomial-1}
\end{figure}

\begin{figure}[t]
\centering
\includegraphics[width=106mm]{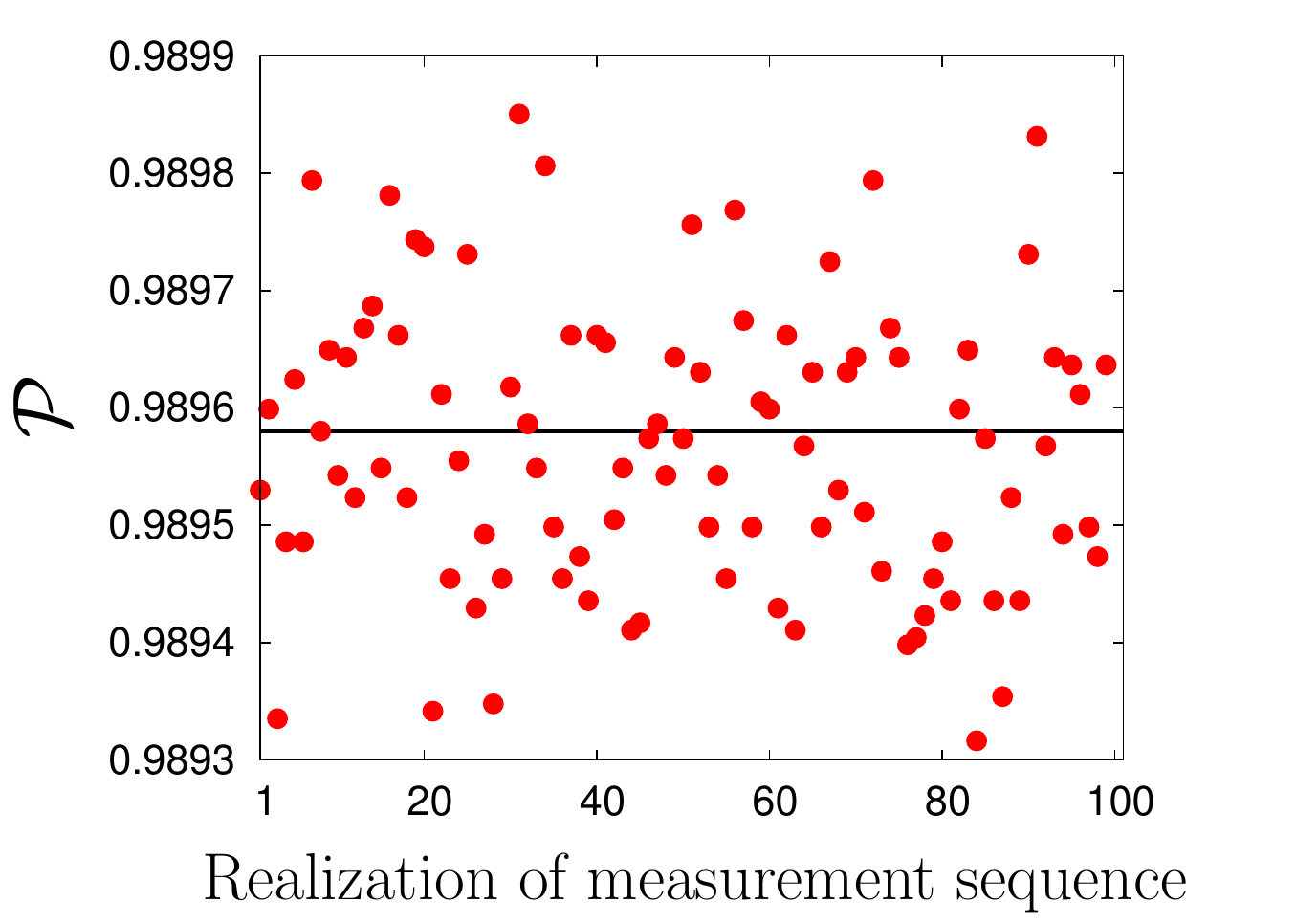}
\caption{Comparison between the survival probability $\mathcal{P}$ obtained by evaluating Eq. (\ref{3}) numerically for $100$ typical
realizations of the measurement sequence (points) and the most probable value $\mathcal{P}^{\star}$ (line) obtained by using Eq. (\ref{20b}), for the case $d=2$
in Fig. \ref{dnomial-1} and for the number of measurements $m=2000$.}
\label{dnomial-2}
\end{figure}

\begin{figure}[b]
\centering
\includegraphics[width=106mm]{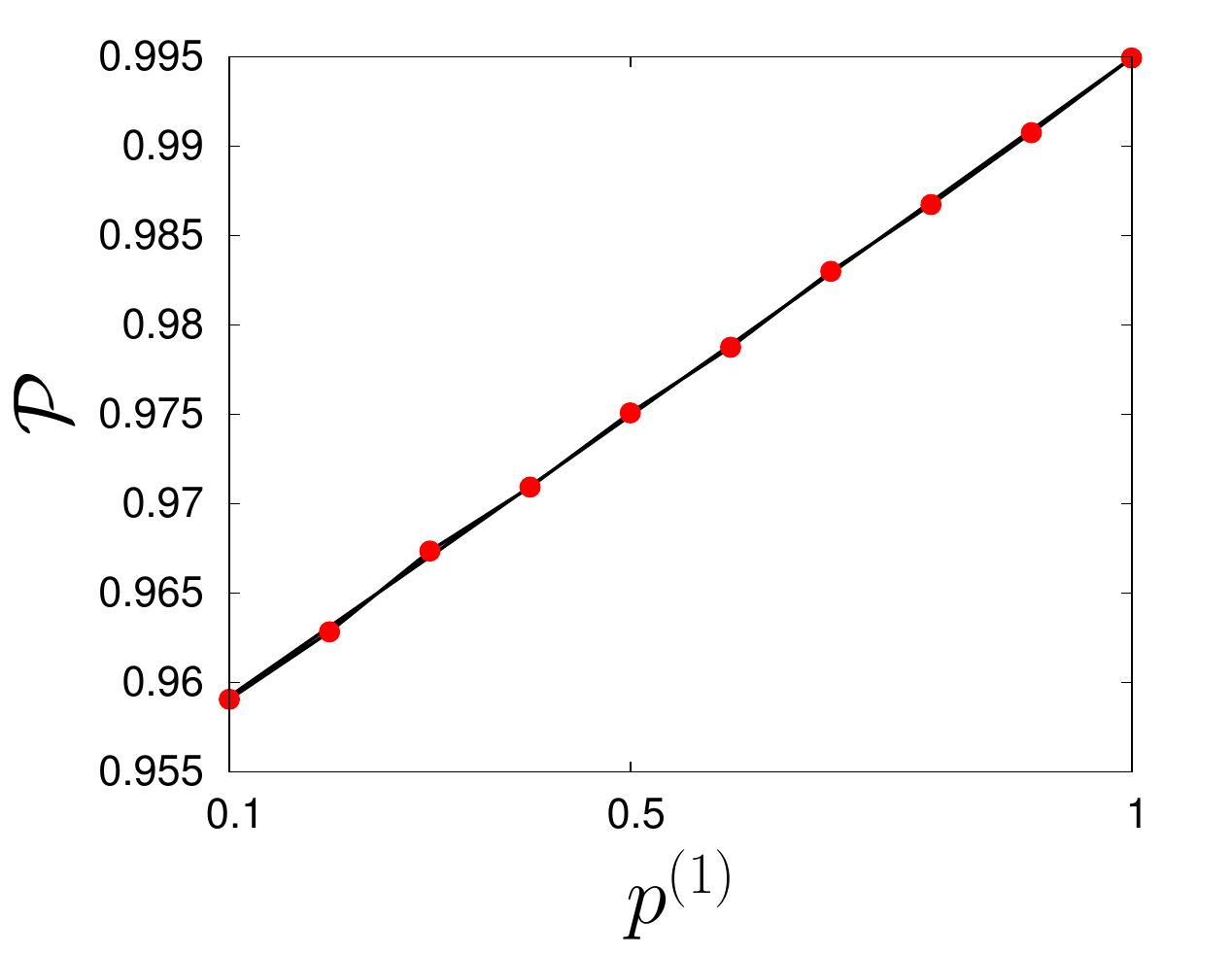}
\caption{Survival probability $\mathcal{P}$ as a function of the probability $p^{(1)}$, with $p^{(2)}=1-p^{(1)}$, for a bi-dimensional Bernoulli
distribution for the $\mu_j$'s and $m=6400$. The points denote the values obtained by evaluating Eq. (\ref{3}) numerically for a typical realization of
the measurement sequence $\{\mu_j\}$, while the line denotes the asymptotic most probable value obtained from Eq.
(\ref{20b}). Here, we have taken $\mu^{(1)}=\mu_0,~\mu^{(2)}=3\mu_0$, with $\mu_0=1$ ns.}
\label{dnomial-3}
\end{figure}

Under these conditions, we obtain the survival probability $\mathcal{P}$ as a function of the number of measurements $m$ for a $d$-dimensional Bernoulli distribution for the $\mu_j$'s,
with $d=2,3,4$ -- see Fig. \ref{dnomial-1}. We find a perfect agreement between the numerical evaluation of Eq. (\ref{3}) for a typical
realization of the measurement sequence $\{\mu_j\}$ and the asymptotic most probable values obtained by using Eq. (\ref{20b}). Moreover, a
comparison between these two quantities for $d=2$, $m=2000$, and $100$ typical realizations of the measurement sequence is shown in Fig. \ref{dnomial-2}.
Note that in the case of a bi-dimensional Bernoulli distribution with probability $p$, the quantity $\mathcal{P}$ depends linearly on $p$ -- see Fig. \ref{dnomial-3}.

\begin{figure}[t!]
\centering
\includegraphics[width=110mm]{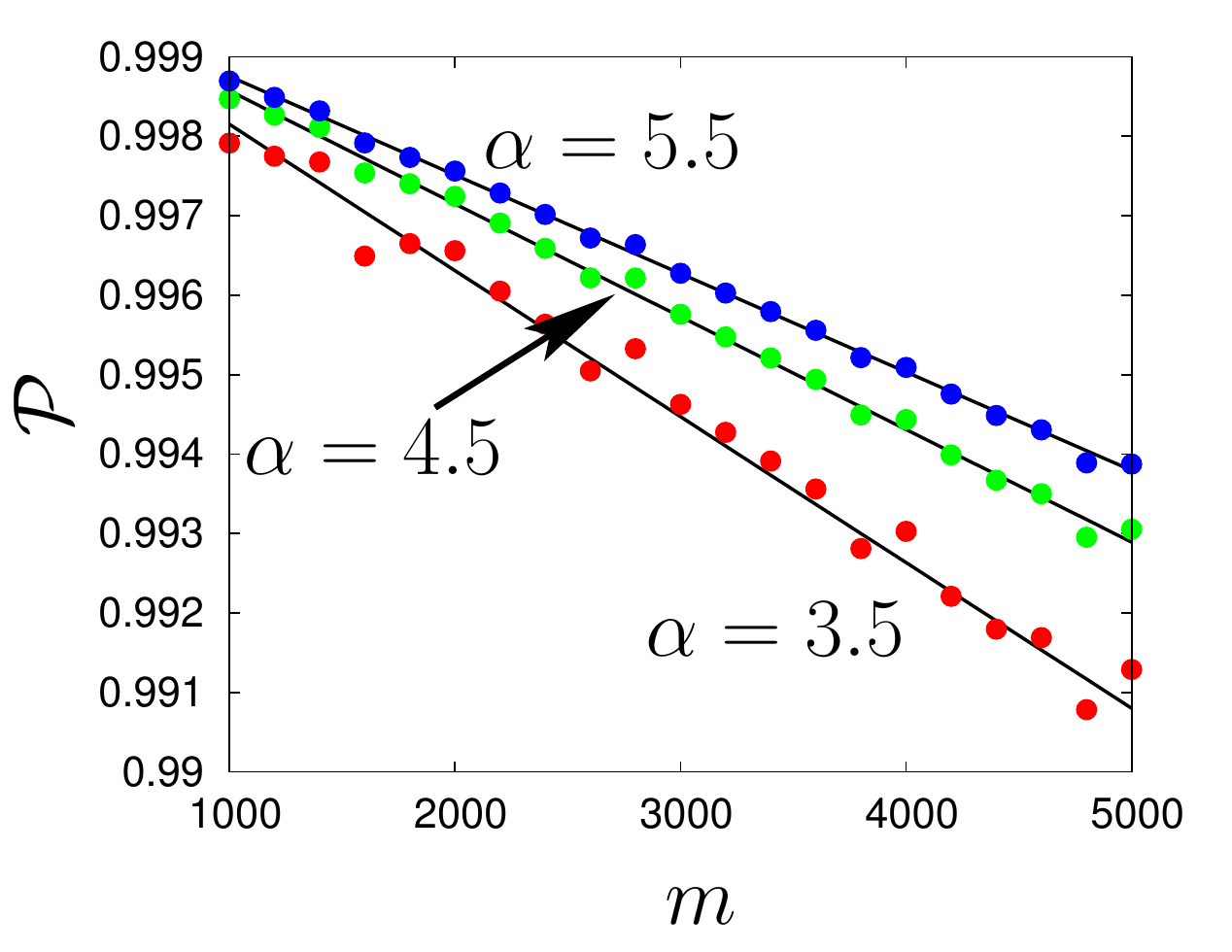}
\caption{Survival probability $\mathcal{P}$ as a function of the number of
measurements $m$ for the distribution
$p(\mu)=\alpha/\Big(\mu_0 (\mu/\mu_0)^{1+\alpha}\Big)$, with $\alpha>0$
and $\mu \in
[\mu_0,\infty]$. Here, $\mu_0$ is a time scale set to $1$ ns. Besides, we choose values of $\alpha$ such that $p(\mu)$ has a finite mean and a finite second moment, i.e. $\alpha > 2$.
As in the preceding figures, the points denote the values obtained by
evaluating numerically Eq. (\ref{3}) for a typical realization of
the measurement sequence $\{\mu_j\}$, while the lines denote the
asymptotic most probable values obtained by using Eq.
(\ref{PG-defn}).}
\label{levy}
\end{figure}

Furthermore, to test our analytical predictions for a continuous $\mu$-distribution, we have considered the following distribution for
the $\mu_j$'s, namely, $p(\mu)=\alpha/\Big(\mu_0 (\mu/\mu_0)^{1+\alpha}\Big)$, with $\alpha>0$ and $\mu \in
[\mu_0,\infty)$. The corresponding survival probability shown in Fig. \ref{levy} further confirms our analytical predictions. Note that the decrease of fluctuations around the most
probable value with increasing $\alpha$ is consistent with the concomitant smaller fluctuations of $\mu$ around the average $\overline{\mu}$. In all the cases discussed here, we observe excellent agreement with our estimate of the most
probable value based on the LD theory. Moreover, our analytical predictions are numerically confirmed also for any coherent
superposition state $|\psi_{0}\rangle \equiv a_1|100\rangle+ a_2|001\rangle$ with $|a_1|^2+|a_2|^2 = 1$. Preliminary numerical
studies show similar results for the numerical survival probability $\mathcal{P}$ in the context of the QZD, by taking into account the
projector $\textbf{P} = |100\rangle \langle 100 | + |001\rangle \langle
001 |$ that confines the system in the Hilbert subspace spanned by the states $|100\rangle$ and $|001\rangle$. It will be further investigated in a forthcoming work.

Finally, we want to address the following question. Does the presence of disorder in the sequence of measurement time intervals enhance the survival probability? To address it, we consider
a $d$-dimensional Bernoulli $p(\mu)$ with $d=2$, and a given fixed value of the average $\overline{\mu}=p^{(1)}\mu^{(1)}+p^{(2)}\mu^{(2)}$. Then, in the first scenario, we apply $m$ projective measurements
at times equally spaced by the amount $\overline{\mu}$, while in the second, we sample this time interval from $p(\mu)$. In the former case, one has
\be\label{pmu-delta}
p(\mu)=\delta(\mu-\overline{\mu}),
\ee
so that Eqs. (\ref{PG-defn}) and (\ref{P-avg1}) give
\begin{equation}\label{QZE-eq0}
\mathcal{P}^\star=\langle \mathcal{P}
\rangle=\exp\Big(m\ln
q(\overline{\mu})\Big)\equiv \mathcal{P}(\overline{\mu}).
\end{equation}
The absence of randomness on the values of $\mu$ trivially leads to $\mathcal{P}^\star=\langle \mathcal{P} \rangle$. In the second scenario, the most probable value $\mathcal{P}^\star$ is given by Eq. (\ref{20b}), hence
\be \label{most-probable-no-random}
\mathcal{P}^\star=\exp\Big(m\Big[p^{(1)}\ln q(\mu^{(1)})+(1-p^{(1)})\ln q(\mu^{(2)})
\Big]\Big); \mu^{(2)}=\frac{\overline{\mu}-p^{(1)}\mu^{(1)}}{p^{(2)}}.
\ee
\begin{figure}[b!]
\centering
\includegraphics[width=115mm]{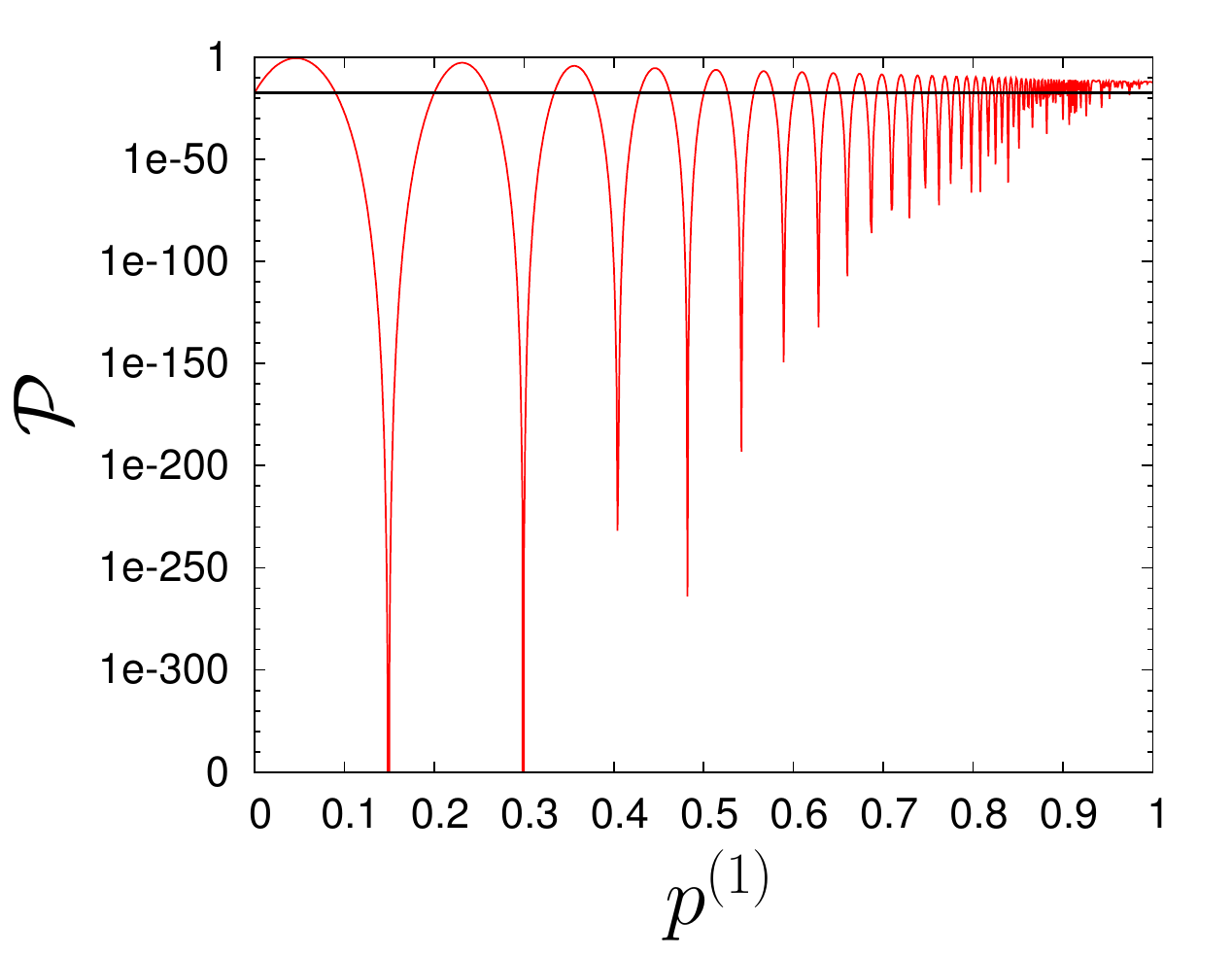}
\caption{Most probable value (red lines) of the survival
probability $\mathcal{P}^{\star}$ in (\ref{most-probable-no-random}) for
a $d$-dimensional Bernoulli distribution $p(\mu)$ with $d=2$, given fixed
values of the average $\overline{\mu}=p^{(1)}\mu^{(1)}+p^{(2)}\mu^{(2)}$ and $\mu^{(1)}$, $m=100$. 
The black line denotes the value $\mathcal{P}^{\star}$ in (\ref{QZE-eq0}) 
in the case of projective measurements equally spaced in time, with
$m=100$. We have considered $\overline{\mu}=2.4\mu_0$, $\mu^{(1)}=\mu_0$, and $\mu_0=10~\mu$s.}
\label{ratio-fig}
\end{figure}
Now the question arises as to whether for given fixed $\overline{\mu}$ and $\mu^{(1)}$ a random sequence of measurement yields
a larger value of the survival probability that the one obtained by performing equally spaced measurements.
For the Hamiltonian (\ref{H}) and the initial state (\ref{initial-state}),
we show in Fig. \ref{ratio-fig} the behavior of $\mathcal{P}^\star$ as a
function of $p^{(1)}$ at fixed values of $\overline{\mu}=2.4\mu_0$ and $\mu^{(1)}=\mu_0$, with $\mu_0=10~\mu$s.
A comparison with $\mathcal{P}(\overline{\mu})$ shows that while in the Zeno limit, such a disorder is deleterious, there are instances where random measurements are beneficial in enhancing the survival probability.
\begin{figure}[h!]
\centering
\includegraphics[scale=0.42]{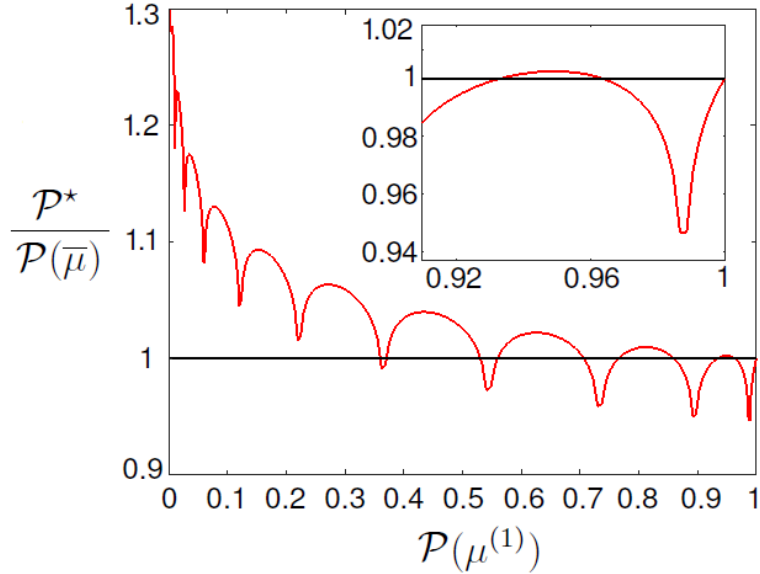}
\caption{Ratio $\mathcal{P}^{\star}/\mathcal{P}(\overline{\mu})$ (red lines) for a d-dimensional Bernoulli distribution $p(\mu)$ with $d=2$ at fixed values of the probability $p^{(1)}
=0.99$, $m=100$ and $\overline{\mu}=2.4\mu^{(1)}$, with $\mu^{(1)}\in[1,250]~ns$. The black line represents the value of $\mathcal{P}^{\star}/\mathcal{P}(\overline{\mu})$ in case of projective measurements equally spaced in time.}
\label{fig6}
\end{figure}
In Fig. \ref{fig6}, moreover, we show the behaviour of the ratio $\mathcal{P}^{\star}/\mathcal{P}(\overline{\mu})$ as a function of $P(\mu^{(1)})$ at fixed values of $p^{(1)}
=0.99$, $m=100$ and $\overline{\mu}=2.4\mu^{(1)}$, with $\mu^{(1)}\in[1,250]~ns$. An effective survival probability enhancement is reached in every dynamical evolution regime, except that in the Zeno limit (inset - Fig. \ref{fig6}).  Interestingly enough, these regimes might be particularly
relevant from the experimental side when the ideal Zeno condition is only partially achieved.

\section{Quantum Zeno limit}
\label{QZE}

Following the analysis in Sec. \ref{LD}, we now discuss how to recover the exact QZ limit. Let us consider $m$ projective measurements at times equally separated by an amount $\overline{\mu}$, so that one has the result (\ref{QZE-eq0}) for the survival probability, while Eq. (\ref{Tavg-1}) gives 
\be
\langle \mathcal{T} \rangle=\mathcal{T}=m\overline{\mu} \; .
\ee
Combining these two expressions, we get
\begin{equation}\label{QZE-eq1}
\mathcal{P}(\overline{\mu})=\exp\Big((\mathcal{T}/\overline{\mu})~\ln
q(\overline{\mu})\Big).
\end{equation}
To recover QZE, in the limit $\overline{\mu}\to 0$ with finite $\mathcal{T}$ and using Eqs. (\ref{qmu-expansion0}) and (\ref{qmu-expansion2}), one indeed obtains
\begin{equation}\label{29}
\mathcal{P}(\overline{\mu})\approx
\exp\Big(-\mathcal{T}\overline{\mu}\Delta^{2}\textbf{H}\Big) \approx 1,
\end{equation}
provided $\Delta^{2}\textbf{H}$ is finite, as it is the case for a
finite-dimensional Hilbert space. 

Let us now discuss QZE for a general $p(\mu)$. Note that in this case, it is natural in experiments to keep the number of measurements $m$ fixed at a large value, with the total time
$\mathcal{T}$ fluctuating between different sequences of measurements $\{\mu_j\}$. From Eqs. (\ref{PG-defn}) and (\ref{P-avg1}), with the use of Eq. (\ref{qmu-expansion0}), and the Taylor expansion of $\log(1+x)$ for $x<1$, we get
\bea \label{Pstar-asymp}
\mathcal{P}^\star&=&\exp\Big(-m\sum_{n=1}^\infty \frac{\langle\delta^n\rangle}{n}\Big)\approx \exp\Big(-m\langle\delta\rangle\Big), \label{QZE-1}\\
\langle \mathcal{P} \rangle&=&\exp\Big(-m \sum_{n=1}^\infty \frac{\langle\delta\rangle^n}{n}\Big)\approx\exp\Big(-m\langle\delta\rangle\Big),\label{QZE-2}
\eea
where
\be
\langle \delta^k \rangle\equiv \int d\mu~p(\mu)\delta^k(\mu);~k=1,2,3,\ldots.
\ee
From Eqs. (\ref{QZE-1}) and (\ref{QZE-2}), it follows that in the limit of very frequent measurements so that $m \to \infty$, provided $\langle \delta \rangle \approx 0$, one recovers the QZE condition, i.e. $\mathcal{P}^\star=\langle\mathcal{P}\rangle \approx 1$. Thus, the condition to obtain QZE in the case of stochastic measurements is
\be
\langle \delta \rangle =\frac{\int d\mu~p(\mu)\mu^2}{\tau_Z^2}\approx 0 \; ,
\ee
which, considering that $\tau_Z$ is finite, reduces to the requirement $\int d\mu~p(\mu)\mu^2 \approx 0$. For instance, for a quite general probability distribution $p(\mu)$ with power-law tails, namely, $p(\mu) \sim 1/(\mu/\mu_0)^{1+\alpha}$, with $~\alpha>0$ and $\mu_0$ being a given time scale, QZE is achieved  for $\mu_0 \ll 1$ and $\alpha>2$ (corresponding to a finite second moment).

\section{Conclusions and outlook}
\label{conclusions}
In this work, we have analyzed stochastic quantum Zeno phenomena by means of the large deviation theory widely applied in probability theory and statistical physics. More
specifically, we have analytically derived the asymptotic distribution of the survival probability for the system to remain in the initial
state when its unitary evolution is combined with a very long sequence of measurements that are randomly spaced in time. The framework allowed
us to obtain analytical expressions for the most probable and the average value of the survival probability. While the most probable value
represents what an experimentalist will measure in a single typical implementation of the measurement sequence, the average value
corresponds instead to an averaging over a large (ideally infinite) number of experimental runs. Therefore, by tuning the probability distribution
of the time intervals between consecutive measurements, one can achieve the most probable survival provability, thereby allowing to engineer
novel optimal control protocols for the manipulation of, e.g., atomic population of a specific quantum state. 

Our analytical predictions have been fully validated against numerical studies of a simple $n$-level quantum system and Zeno-protected entangled states. These states are
particularly relevant in the context of quantum information science, and therefore need to be well protected from unavoidable environmental
decoherence \cite{Wang1,Maniscalco1}. For example, in Ref. \cite{zhang}, it has been shown that an entangled state can be characterized by a Zeno
time comparable with the one of a separable state by virtue of its interaction with a suitable noisy environment. Since the decoherence may
correspond to a continuous monitoring from the environment (repetitive random measurements), our formalism allows one to predict the occupation
probability of an arbitrary entangled state by the knowledge of the probability distribution of the system-environment interaction times.
Additionally, we have found that the presence of stochasticity in the measurement process may enhance the survival probability in the typical
experimental scenario where quantum Zeno limit is not achieved. This suggests new schemes that advantageously exploits disorder in
implementing quantum information protocols.

Finally, as an outlook, our formalism may be extended to the more general context of QZD where the system dynamics gets confined in a
Hilbert space, thereby enhancing its complexity exponentially \cite{Giovannetti1}. Another possible related application is the analysis of the interplay 
between classical noise arising from external drive and the noise due to the randomness in the measurement sequence, thereby designing new types of noise constraining the Hamiltonian 
dynamics in atomic quantum simulators of many-body systems \cite{Stannigel1}.

\section{Acknowledgments}
We acknowledge fruitful discussions with M. Campisi and H. Touchette. F.S.C. acknowledges support of MIUR (Project No.
2010LLKJBX). The work of F.C. is supported by EU FP7 Marie-Curie Programme (Career Integration Grant No. 293449) and by a MIUR-FIRB grant (Project No. RB FR10M3SB).
\appendix

\section{Derivation of Eq. (\ref{15})}
\label{app1}
In this Appendix, we derive Eq. (\ref{15}) of the main text. From Eqs.
(\ref{12}) and (\ref{12a}), we get
\begin{equation}\label{appeq1}
m \ln
q\left(\mu^{(d)}\right)-\mathcal{L}=\sum_{\alpha=1}^{d-1}n_{\alpha}\lambda\left(\mu^{(\alpha)}\right),
\end{equation}
with
\begin{equation}
\lambda(\mu^{(\alpha)})\equiv\ln q(\mu^{(d)})- \ln q(\mu^{(\alpha)}).
\end{equation}
Equation (\ref{appeq1}) is solved with
\begin{equation}\label{sol1}
n_{\alpha}'=\frac{m\ln q\left(\mu^{(d)}\right)-\mathcal{L}}
{(d-1)\lambda\left(\mu^{(\alpha)}\right)};~\alpha=1,2,\ldots,d-1,
\end{equation}
while $n_d'$ is given by
\be\label{app0-eq1}
n_d'=m-\sum_{\alpha=1}^{d-1}n_{\alpha}'.
\ee
Then, Eq. (\ref{eq1}) gives 
\begin{eqnarray}
P(\mathcal{L})&=&\exp\left\{\ln m!-\sum_{\alpha=1}^{d}\ln n_{\alpha}'!+
\sum_{\alpha=1}^{d}n_{\alpha}'\ln p^{(\alpha)}\right\} \nonumber \\
&\approx&\exp\left\{m\ln m -m-\sum_{\alpha=1}^{d}n_{\alpha}'\ln n_{\alpha}'+\sum_{\alpha=1}^{d}n_{\alpha}'+
\sum_{\alpha=1}^{d}n_{\alpha}'\ln p^{(\alpha)}\right\} \nonumber \\
&=&\exp\left\{m\ln m -m\sum_{\alpha=1}^{d-1}\frac{\ln q\left(\mu^{(d)}\right)-\frac{\mathcal{L}}{m}}{(d-1)\lambda\left(\mu^{(\alpha)}\right)}\ln\left(m\frac{\ln q\left(\mu^{(d)}\right)-\frac{\mathcal{L}}{m}}{(d-1)\lambda\left(\mu^{(\alpha)}\right)}\right)\right.\nonumber \\
&&\left.-m\left(1-\sum_{\alpha=1}^{d-1}
\frac{\ln q\left(\mu^{(d)}\right)-\frac{\mathcal{L}}{m}}{(d-1)\lambda\left(\mu^{(\alpha)}\right)}\right)\ln\left[m\left(1-\sum_{\alpha=1}^{d-1}\frac{\ln q\left(\mu^{(d)}\right)-\frac{\mathcal{L}}{m}}{(d-1)\lambda\left(\mu^{(\alpha)}\right)}\right)\right]\right.\nonumber \\
&&\left.+m\sum_{\alpha=1}^{d-1}\frac{\ln q\left(\mu^{(d)}\right)-\frac{\mathcal{L}}{m}}{(d-1)\lambda\left(\mu^{(\alpha)}\right)}\ln p^{(\alpha)}\right.\nonumber \\
&&\left.+m\left(1-\sum_{\alpha=1}^{d-1}
\frac{\ln q\left(\mu^{(d)}\right)-\frac{\mathcal{L}}{m}}{(d-1)\lambda\left(\mu^{(\alpha)}\right)}\right)\ln\left(1-\sum_{\alpha=1}^{d-1}p^{(\alpha)}\right)\right\}\nonumber\\
&\approx&\exp\Big[-mI\left(\frac{\mathcal{L}}{m}\right)\Big]
\label{app0-eq2},
\end{eqnarray}
where, in the second step, we have used the Stirling's approximation,
while in the third step we have used Eqs. (\ref{sol1}) and
(\ref{app0-eq1}). Here, we have
\begin{eqnarray}
&&I\left(x\right)=\sum_{\alpha=1}^{d}f(\mu^{(\alpha)})\ln\left(\frac{f(\mu^{(\alpha)})}{p^{(\alpha)}}\right),
\\
&&f(\mu^{(\alpha)})=\frac{\ln q(\mu^{(d)})-x}{(d-1)\lambda(\mu^{(\alpha)})}\label{appeq2};~\alpha=1,\ldots,(d-1), \\
&&f(\mu^{(d)})=1-\sum_{\alpha=1}^{d-1}f(\mu^{(\alpha)}).
\end{eqnarray}
Equation (\ref{app0-eq2}) is Eq. (\ref{15}) of the main text.

\section{Derivation of Eq. (\ref{21})}
\label{app2}
Here, we provide more details on the derivation of Eq. (\ref{21}). From Eq. (\ref{16}), the condition $\partial
I\left(\mathcal{L}/m\right)/\partial \ln
q(\mu^{(\alpha)})|_{\mathcal{L}=\mathcal{L}^\star}=0$ gives for
$\alpha=1,\ldots,d-1$ the relation
\begin{equation}\label{appeq3}
p^{(d)}f(\mu^{(\alpha)})=p^{(\alpha)}\left(1-\sum_{\alpha=1}^{d-1}f(\mu^{(\alpha)})\right).
\end{equation}
Summing both sides over $\alpha=1,2,\ldots,d-1$, we get
\be
p^{(d)}\sum_{\alpha=1}^{d-1}f(\mu^{(\alpha)})=\left(1-\sum_{\alpha=1}^{d-1}f(\mu^{(\alpha)})\right)\sum_{\alpha=1}^{d-1}p^{(\alpha)},
\ee
which, by using $\sum_{\alpha=1}^dp^{(\alpha)}=1$, gives
\be
\sum_{\alpha=1}^{d-1}f(\mu^{(\alpha)})=\sum_{\alpha=1}^{d-1}p^{(\alpha)}.
\ee
Using the above equation, and combining Eqs. (\ref{appeq2}) and (\ref{appeq3}), one has
\begin{equation}
\left(\ln
q(\mu^{(d)})-\frac{\mathcal{L}^\star}{m}\right)=(d-1)\frac{\left(1-\displaystyle{\sum_{\alpha=1}^{d-1}}p^{(\alpha)}\right)}{p^{(d)}}p^{(\alpha)}\lambda(\mu^{(\alpha)}),
\end{equation}
that is,
\be
\frac{\mathcal{L}^\star}{m}=\ln
q(\mu^{(d)})-\left(1-\sum_{\alpha=1}^{d-1}p^{(\alpha)}\right)\left(\sum_{\alpha=1}^{d-1}
\lambda(\mu^{(\alpha)})\frac{p^{(\alpha)}}{p^{(d)}}\right),
\ee
yielding finally
\be
\mathcal{L}^\star=m\sum_{\alpha=1}^{d} p^{(\alpha)}\ln
q(\mu^{(\alpha)}),
\ee
which is Eq. (\ref{21}) of the main text.

\section{Derivation of Eq. (\ref{LD-PLT})}
\label{app3}
To derive Eq. (\ref{LD-PLT}), we use Eqs. (\ref{cons3}) and (\ref{T-defn1}) to get
\be
\sum_{\alpha=1}^d n_\alpha' \mu^{(\alpha)}=m\overline{\mu},
\ee
which, by rewriting Eq. (\ref{cons1}) as
\be
(m-\sum_{\alpha=1}^{d-1}n_{\alpha}')\mu^{(d)}+\sum_{\alpha=1}^{d-1}
n_\alpha' \mu^{(\alpha)}=m\overline{\mu},
\ee
leads to
\begin{equation}\label{app-eq31}
m=\frac{\displaystyle{\sum_{\alpha=1}^{d-1}n_{\alpha}'\left(\mu^{(d)}-\mu^{(\alpha)}\right)}}{(\mu^{(d)}-\overline{\mu})}.
\end{equation}

Equations (\ref{cons1}) and (\ref{cons2}) then give (see the derivation of Eq. (\ref{sol1}))
\begin{equation}\label{app-eq32}
n_{\alpha}'=\frac{m\ln q\left(\mu^{(d)}\right)-\mathcal{L}}
{(d-1)\lambda\left(\mu^{(\alpha)}\right)};~\alpha=1,2,\ldots,d-1,
\end{equation}
while $n_d'$ is
\be
n_d'=m-\sum_{\alpha=1}^{d-1}n_{\alpha}'.
\ee
Combining Eqs. (\ref{app-eq31}) and (\ref{app-eq32}), and noting that
$m\ne 0$, we get
\begin{equation}\label{d12}
\sum_{\alpha=1}^{d-1}\frac{\left(\ln
q\left(\mu^{(d)}\right)-\mathcal{L}/m\right)(\mu^{(d)}-\mu^{(\alpha)})}{(d-1)\lambda\left(\mu^{(\alpha)}\right)(\mu^{(d)}-\overline{\mu})}=1,
\end{equation}
which is satisfied with $\left(\ln
q\left(\mu^{(d)}\right)-\mathcal{L}/m\right)(\mu^{(d)}-\mu^{(\alpha)})=(d-1)\lambda\left(\mu^{(\alpha)}\right)(\mu^{(d)}-\overline{\mu})~~\forall~~\alpha=1,\ldots,(d-1)$.

From Eq. (\ref{app-eq32}), we get for $\alpha=1,2,\ldots,d-1$ that
\bea
n_\alpha'&=&\frac{\mu^{(d)}(m\ln q\left(\mu^{(d)}\right)-\mathcal{L})}
{(d-1)\lambda\left(\mu^{(\alpha)}\right)(\mu^{(d)}-\overline{\mu}+\overline{\mu})}\\
&=&\frac{\mu^{(d)}(m\ln q\left(\mu^{(d)}\right)-\mathcal{L})}
{\left(\ln
q\left(\mu^{(d)}\right)-\mathcal{L}/m\right)(\mu^{(d)}-\mu^{(\alpha)})+(\mathcal{T}/m)(d-1)\lambda\left(\mu^{(\alpha)}\right)}.
\eea
Using the last expression, and proceeding in a way similar to
\ref{app1}, we get
\begin{eqnarray}\label{app3-eq1}
P(\mathcal{L},\mathcal{T})&\approx&\exp\Big(-m\mathcal{I}\left(\frac{\mathcal{L}}{m},\frac{\mathcal{T}}{m}\right)\Big),
\end{eqnarray}
where
\begin{eqnarray}
&&\mathcal{I}\left(x,y\right)=\sum_{\alpha=1}^{d}g(\mu^{(\alpha)})\ln\left(\frac{g(\mu^{(\alpha)})}{p^{(\alpha)}}\right),
\\
&&g(\mu^{(\alpha)})=\frac{\mu^{(d)}(m\ln q\left(\mu^{(d)}\right)-x)}
{\left(\ln
q\left(\mu^{(d)}\right)-x\right)(\mu^{(d)}-\mu^{(\alpha)})+y(d-1)\lambda\left(\mu^{(\alpha)}\right)};~\alpha=1,\ldots,(d-1),
\nonumber \\ \\
&&g(\mu^{(d)})=1-\sum_{\alpha=1}^{d-1}g(\mu^{(\alpha)}).
\end{eqnarray}
Equation (\ref{app3-eq1}) is Eq. (\ref{LD-PLT}) of the main text.

\end{document}